\documentclass[aps,prb,twocolumn,superscriptaddress,floatfix,longbibliography]{revtex4-2}

\usepackage{amsmath,amssymb} 
\usepackage{bm} 
\usepackage{graphicx} 
\usepackage{comment} 
\usepackage{textcomp} 
\usepackage{gensymb} 
\usepackage{ulem} 
\usepackage{booktabs}
\usepackage{makecell}
\usepackage{float}

\usepackage{xr-hyper}
\makeatletter
\newcommand*{\addFileDependency}[1]{
  \typeout{(#1)}
  \@addtofilelist{#1}
  \IfFileExists{#1}{}{\typeout{No file #1.}}
}
\makeatother

\usepackage{dcolumn} 
\usepackage{bm}     
\usepackage{amsfonts}
\usepackage[bookmarks=false,colorlinks,citecolor=blue]{hyperref}
\hypersetup{colorlinks=true,linkcolor=blue,filecolor=blue,citecolor = blue, urlcolor=blue}\usepackage{amsmath}
\newcommand{\angstrom}{\textup{\AA}}
\usepackage[version=4]{mhchem}
\usepackage{siunitx}
\usepackage{xcolor}
\usepackage{textcomp}

\usepackage{mathtools}

\DeclarePairedDelimiterX\braket[2]{\langle}{\rangle}{#1\,\delimsize\vert\,\mathopen{}#2}

\usepackage{enumitem}
\setlist{noitemsep,leftmargin=*,topsep=0pt,parsep=0pt}

\usepackage{xcolor} 
\definecolor{lightgray}{gray}{0.6}
\definecolor{medgray}{gray}{0.4}

\newif\ifptitle
\newif\ifpnumber
\newcounter{para}

\ptitletrue  
\pnumbertrue  

\newcommand{\mytitle}{Heterogeneous Transfer of Thin Film BaTiO$_3$ onto Silicon for Device Fabrication}

\begin{document}

\title{\mytitle}

\author{Temazulu S. Zulu}
\affiliation{Department of Physics, Harvard University, Cambridge, MA, USA}

\author{Larissa B. Little}
\affiliation{Department of Physics, Harvard University, Cambridge, MA, USA}
\affiliation{School of Engineering and Applied Science, Harvard University, Cambridge, MA, USA}

\author{Aaron M. Day}
\affiliation{Department of Physics, Harvard University, Cambridge, MA, USA}
\affiliation{School of Engineering and Applied Science, Harvard University, Cambridge, MA, USA}
\affiliation{Harvard Quantum Initiative, Harvard University, Cambridge, MA, USA}

\author{Chaoshen Zhang}
\affiliation{School of Engineering and Applied Science, Harvard University, Cambridge, MA, USA}

\author{Keith Powell}
\affiliation{School of Engineering and Applied Science, Harvard University, Cambridge, MA, USA}

\author{Kyeong-Yoon Baek}
\affiliation{Department of Physics, Harvard University, Cambridge, MA, USA}

\author{Benazir Fazlioglu-Yalcin}
\affiliation{Department of Physics, Harvard University, Cambridge, MA, USA}

\author{Neil Sinclair}
\affiliation{School of Engineering and Applied Science, Harvard University, Cambridge, MA, USA}

\author{Charles M. Brooks}
\affiliation{Department of Physics, Harvard University, Cambridge, MA, USA}

\author{David R. Barton}
\affiliation{Department of Materials Science and Engineering, Northwestern University, Evanston, IL, USA}

\author{Marko Lon\v{c}ar}
\affiliation{School of Engineering and Applied Science, Harvard University, Cambridge, MA, USA}

\author{Julia A. Mundy}
\thanks{\href{mailto:mundy@fas.harvard.edu}{mundy@fas.harvard.edu}}
\affiliation{Department of Physics, Harvard University, Cambridge, MA, USA}
\affiliation{School of Engineering and Applied Science, Harvard University, Cambridge, MA, USA}
\date{\today}

\newcommand{\BTO}{BaTiO$_3$}

\begin{abstract}

Thin film \BTO~has one of the highest known Pockels coefficients ($>$1200 pm/V), making it an attractive material for use in electro-optic devices. It is advantageous to integrate \BTO~on silicon to enable complementary metal-oxide-semiconductor (CMOS) compatible processing. However, synthesis of high-quality \BTO~directly on silicon remains a challenge. Here, we synthesize \BTO~using hybrid metal-organic molecular beam epitaxy (hMBE) and demonstrate its transfer onto silicon using thermocompression bonding and chemical lift-off. Hybrid metal-organic MBE enables self-regulated synthesis of highly stoichiometric thin films at high growth rates ($>$100nm/hr). Our transfer method results in millimeter-scale areas of atomically flat, crack-free \BTO~making it a potentially scalable method. Finally, we demonstrate the applicability of our process to device fabrication through characterization of lithographically-patterned and etch-transferred sub-micron features.

\end{abstract}
\maketitle

\section{Introduction}

Thin film barium titanate (\BTO) (see Fig. 1A) has emerged as a leading candidate for integrated photonics due to its extraordinary electro-optical properties. \BTO~has a Pockels coefficient of over 1200 pm/V in thin film form, which is about 40 times higher than the leading material platform for electro-optic (EO) integrated photonics, thin film lithium niobate ({LiNbO$_3$}) \cite{lin_giant_2025, wen_fabrication_2024}. \BTO~ also has a high refractive index ($\sim{2.5}$), a wide band gap ($\sim{3.2}$ eV)  and has low losses at visible and near infrared wavelengths \cite{karvounis_barium_2020}. Indeed, \BTO~based photonic devices have been demonstrated with low losses (below 0.15 dB/cm) and quality factors of above 1 million \cite{raju_high-q_2025, kim_low_2025}. A key to achieving these results in devices is to have \BTO~integrated onto silicon for device fabrication, enabling Complementary Metal-Oxide-Semiconductor (CMOS) compatible processing and characterization \cite{xiong_active_2014, abel_large_2019}. 

While \BTO~is a promising electro-optic material, challenges in synthesis have impeded its wide-spread adoption. LiNbO$_3$ can be synthesized as large single crystals, which can be made into thin films using smart-cut technology \cite{hu_integrated_2025, yang_advanced_2025}. This results in LiNbO$_3$ on SiO$_2$/Si, enabling effective light confinement with the low index of refraction SiO$_2$ layer adjacent to the electro-optic LiNbO$_3$. 
In contrast, high quality \BTO~ cannot be synthesized as large single crystals. \BTO~thin films have been synthesized on silicon with thin seed layers of SrTiO$_3$ to alleviate the chemical incompatibility of the \BTO/Si interface \cite{rahim_taking_2021}. This stack is then bonded to SiO$_2$/Si and the back silicon wafer etched \cite{abel_large_2019}. However, synthesizing highly crystalline \BTO/SrTiO$_3$ on silicon continues to be challenging; in addition to the epitaxial strain, SrTiO$_3$ has also been reported to absorb hydrogen during the fabrication process, leading to propagation losses in \BTO~devices (6 Db/cm) \cite{rahim_taking_2021,eltes_low-loss_2016}.

To avoid the challenges of synthesis on silicon, \BTO~can instead be synthesized on iso-structural substrates and then transferred onto SiO$_2$/Si using different lift-off methods. For example, BaTiO$_3$ has been synthesized on graphene-coated perovskite substrates. Exploiting the van der Waals bonding of graphene and the substrate, the film can be exfoliated as a freestanding membrane and then transferred \cite{yoon_freestanding_2022,kum_heterogeneous_2020}. While this is a powerful technique, it is challenging to obtain a coherent film \cite{choo_oxide_2024}, especially due to the highly oxidizing conditions required for \BTO~synthesis \cite{yoon_freestanding_2022}.  Alternatively, recent work has demonstrated that BaTiO$_3$ can be synthesized on soluble buffer layers (chemical lift-off). The \BTO~thin films are subsequently coated with polymer supporting layers and the buffer layer dissolved to form a membrane that is transferred to silicon \cite{cao_transfer-heterogeneous_2025,wang_towards_2020, lee_ultrathin_2022}.
However, polymer residue degrades the quality of \BTO~\cite{chiabrera_freestanding_2022,liang_toward_2011}. 

Here, we report the transfer of millimeter-scale flat, crack-free BaTiO$_3$ onto SiO$_2$/Si using a combination of thermocompression bonding and chemical lift-off. We synthesize epitaxial \BTO~on water soluble strontium oxide (SrO) sacrificial layers using hybrid metal-organic molecular beam epitaxy (hMBE; see Fig. 1B). Hybrid metal-organic MBE enables the deposition of stoichiometric thin films at high growth rates ($>$100 nm/hr) over a large self-regulated synthesis window, providing a reproducible and scalable method for synthesis of \BTO~for device fabrication \cite{fazlioglu-yalcin_stoichiometric_2023, nunn_hybrid_2021, lapano_scaling_2019}. Chemical lift-off has recently been demonstrated to have the potential to have large freestanding membranes \cite{lu_synthesis_2016,zhang_super-tetragonal_2024,dong_super-elastic_2019,chiabrera_freestanding_2022}, however, millimeter scale films are prone to cracking and wrinkling, reducing the functional area for device fabrication. Combining chemical lift-off and thermocompression bonding avoids the use of polymer supporting layers and freestanding membranes, reducing residue and membrane cracking. Thermocompression bonding, compatible with existing industry processes, provides stress during transfer, ensuring that the resulting film is flat and adhered to silicon \cite{bouaziz_advanced_2023}. We characterize transferred thin films on silicon and perform lithography and etch tests to demonstrate the robustness to nanofabrication processes. Our films feature low surface roughness and our transfer method is robust to monolithic etching of sub-micron sized features. We thus develop a potentially scalable method towards large area \BTO~thin films on silicon for electro-optic device fabrication.

\section{Results and Discussion}
\begin{figure}[h]
    \includegraphics[width = \columnwidth]{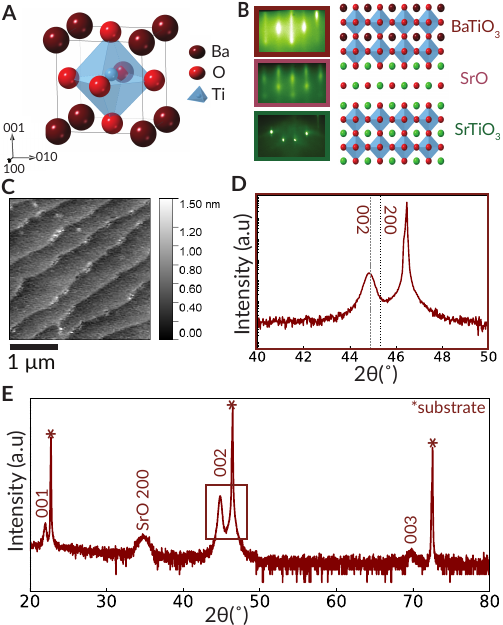}
    \caption{\textit{Synthesis and characterization of thin film \BTO}. \textbf{A.} The tetragonal structure of a \BTO~unit cell. The central atom, titanium, is slightly displaced in the $c$-axis direction, resulting in a ferroelectric polarization. \textbf{B.} We synthesize \BTO~on (100)-orientated {SrTiO$_3$} substrates with an intermediate sacrificial SrO layer (thickness not to scale). \textit{In situ} reflection high energy electron diffraction (RHEED) patterns are shown for each step of the deposition, along the [100] substrate azimuth. \textbf{C.} AFM image of the \BTO~thin film on SrO/SrTiO$_3$ displays atomic step terraces and a root mean squared roughness (RMS) of 153.3 picometers. \textbf{D.} The \BTO~002 peak in XRD indicates a $c$-axis oriented film. \textbf{D.} Full XRD pattern of \BTO~grown on an SrO sacrificial layer and SrTiO$_3$ substrate.}
     \label{fig:one}
\end{figure}

\subsection{Synthesizing high quality \BTO}
We used hMBE to synthesize SrO sacrificial layers and epitaxial \BTO~thin films. The SrO and \BTO~layers grow epitaxially as shown in Fig. 1B. Using \textit{in situ} reflection high energy electron diffraction (RHEED), we are able to determine the surface morphology of the thin films during synthesis. The RHEED image following the SrO deposition shows modulated streaks, indicating a multilevel stepped surface (Fig. 1B) \cite{hasegawa_reflection_2012} as a result of the +7$\%$ lattice mismatch between the {SrTiO$_3$} and SrO. Nevertheless, the RHEED image after the BaTiO$_3$ deposition (Fig. 1B) shows narrow streaks that signify a flat film surface with small domains \cite{hasegawa_reflection_2012}. Post synthesis, atomic-force microscopy (AFM) imaging displays a smooth thin film of \BTO~on SrO with atomic step terraces and a root mean square roughness of 153.3 picometers (Fig. 1C). 

 Figure 1D-E display the x-ray diffraction (XRD) pattern of the 10 nm SrO and 100 nm \BTO~synthesized on (100)-orientated {SrTiO$_3$}. A thin SrO layer minimized degredation in air (See Supp. Fig. S2) \cite{varshney_hybrid_2024-1}. We obtain single crystalline SrO with a lattice parameter of 5.162 $\si{\angstrom}$ using Nelson-Riley fits \cite{nelson_experimental_1945}; this is close to the reported bulk crystal value of 5.159 $\si{\angstrom}$, with a slight deviation that could be attributed to tensile strain. The \BTO~thin film is single crystalline with the $c$-axis oriented in the out-of-plane direction. We extract a $c$-axis lattice parameter of 4.039 $\si{\angstrom}$ for the BaTiO$_3$ film.  
 The relaxation of \BTO~and SrO layers synthesized by hMBE is consistent with previous findings by other groups \cite{varshney_hybrid_2024, choo_hybrid_2025, fazlioglu-yalcin_stoichiometric_2023}.
 
\subsection{Transferring \BTO~onto silicon}

Following the synthesis of high quality BaTiO$_3$ on SrO/SrTiO$_3$, we demonstrate the transfer of millimeter sized flat, crack-free \BTO~onto SiO$_2$/Si. As shown schematically in Fig. 2, we first deposit {SiO$_2$} on \BTO~and on silicon substrate wafers using chemical vapor deposition (CVD, see Methods). We then deposit gold, using electron beam evaporation, on the {SiO$_2$} using chromium film as an adhesion layer. Finally, we use thermocompression bonding to adhere the two gold surfaces to each other. We then place the stack in DI water to dissolve the SrO, releasing the BaTiO$_3$ layer from the initial SrTiO$_3$ substrate, constructing a thin film of \BTO~that is bonded to a layer of SiO$_2$ and adhered to a silicon wafer.  We explored \BTO~films of various thicknesses ranging from 10 nm to 100 nm; we found that thicker films led to less cracking and larger regions of flat transfers onto silicon (see Suppl. Fig. S1). All the subsequent results shown are based on 100 nm \BTO~transferred onto silicon.

\begin{figure*}
    \includegraphics[width = 2\columnwidth]{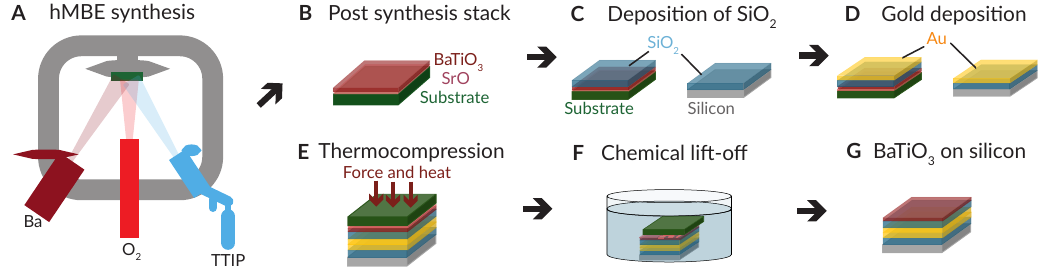}
    \caption{\textit{Transfer  of \BTO~onto silicon}. \textbf{A.} We use hMBE to deposit strontium oxide (SrO) sacrificial layers and thin film \BTO. \textbf{B.} The post synthesis stack with \BTO~on SrO. \textbf{C.} {SiO$_2$} is deposited onto the \BTO~thin film and on a silicon substrate using CVD. \textbf{D.} Gold is deposited on top of the {SiO$_2$} layers using electron beam evaporation. \textbf{E.} Thermocompression bonding with a force of 1000 N and 220\textdegree C is performed on the gold surfaces of \BTO~and silicon. \textbf{F.} The stack is immersed in deionized (DI) water to dissolve the SrO sacrificial layer. \textbf{G.} The final stack has \BTO~on a SiO$_2$ layer on a silicon wafer.}
     \label{fig:two}
\end{figure*}

Following the transfer onto silicon, we perform XRD scans of \BTO~. We find the \BTO~film is $c$-axis oriented with Bragg reflections demonstrating the preservation of a single crystalline thin film structure (Fig. 3A). In Fig. 3B, we compare the \BTO~002 peak on the as synthesized sample (in maroon) and the transferred sample (in pink) and we observe a slight shift in the $c$-axis lattice parameter to 4.035 $\si{\angstrom}$  after transfer; this is consistent with the expected bulk $c$-axis lattice parameter of 4.036 $\si{\angstrom}$. This shift could be attributed to a small strain during the thermal compression bonding due to the differing thermal expansion coefficients or a slight degradation during the dissolution process (See Suppl. Fig. S1, Suppl. Fig. S2). Reciprocal space mapping (RSM, Fig. 3C) shows that the \BTO~302 peak is relaxed.
  
The AFM image of \BTO~on silicon is displayed in Fig. 3D, demonstrating atomic step terraces and a root mean squared roughness of 369 picometers. The presence of atomic step terraces has also been seen in freestanding membrane transfers \cite{wohlgemuth_chemical_2025, yun_strain_2024}. Notably, there is some residue on the edges of the step terraces that could possibly be reduced with further cleaning steps (DI water sonication, acetone sonication, IPA sonication).  Finally, the Scanning electron microscopy (SEM) image illustrates a millimeter sized region of flat, crack-free \BTO~on a silicon substrate (Fig. 3E). 

\begin{figure}
    \includegraphics[width = \columnwidth]{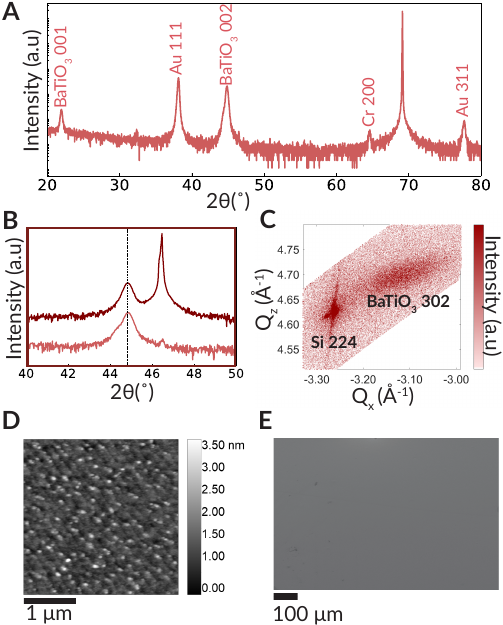}
    \caption{\textit{\BTO~transferred onto silicon}. \textbf{A.} The XRD pattern of \BTO~ transferred onto a silicon substrate with the out-of-plane Bragg reflections (00$\textit{l}$) of \BTO~showing that we have preserved the single crystalline structure after transfer onto silicon.  \textbf{B.} XRD pattern of the 002 \BTO~peak before transfer (maroon) and after transfer (pink). We note that there is a very slight shift in the peaks due to compressive strain from the bonding stack. \textbf{C.} The reciprocal space map (RSM) of \BTO~ 302 peak on silicon 103 peak shows a relaxed thin film in relation to the substrate. \textbf{D.} The AFM image of \BTO~on silicon depicts atomic step terraces with a root mean squared roughness of 369 picometers.  \textbf{E.} The scanning electron microscopy (SEM) image of \BTO~on silicon shows a millimeter scale region that is flat and crack-free.} 
     \label{fig:three}
\end{figure}

\subsection{Etch testing on transferred \BTO}

Finally, we demonstrate the applicability of our transfer process to device fabrication by performing lithography and dry etch tests. We pattern an array of nanopillars with feature sizes down to 1 $\mu$m onto the \BTO~on silicon using electron beam lithography. We then perform reactive ion dry etching and show via SEM etched pillars (Fig. 4). The etched features remain intact over hundred micron areas, even after a resist removal process, indicating a robust thin film.

\begin{figure}
    \includegraphics[width = 1\columnwidth]{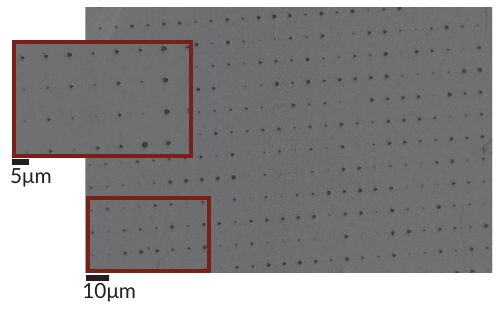}
    \caption{\textit{\BTO~fabrication tests}.
    The SEM of nanopillars with feature sizes down to 1 $\mu$m is shown, demonstrating the \BTO~is robust to standard nanofabrication.}
     \label{fig:four}
\end{figure}

\section{Conclusions}

 Here, we demonstrate the synthesis of single crystalline \BTO~on iso-structural and nearly lattice matched substrates. By combining thermocompression bonding and chemical lift-off, we transfer millimeter scale areas of flat, crack-free \BTO~onto silicon. While thermocompression bonding and chemical lift-off had previously been demonstrated as individual techniques, they had substantial trade-offs: wafer bonding had been used for lower quality \BTO~synthesized on heterogeneous silicon wafers and chemical lift-off had been used to form freestanding membranes that were prone to cracking or wrinkling. Here we are able to perform bonds at low temperatures and have films flatly adhered onto silicon while avoiding cracking, wrinkling and residues from chemical-lift off on its own. 
Our new method can be implemented with existing fabrication processes with a short turnaround time.
 
 Our results are a step towards mitigating the synthesis and transfer challenges for \BTO~integration into electro-optic device fabrication. Future work could optimize this technique for fully wafer-scale transfer. Our work used thermocompression bonding outside of a cleanroom environment. Future work could explore this process within a cleanroom which could enable even larger scale transfer with ultraclean gold surfaces. While our technique demonstrated the heterogenous transfer from an oxide to a silicon substrate, a closer coefficient of thermal expansion between the film and substrate could lead to transfer which preserves the precise lattice constant of the as-synthesized film.   Having freedom in substrate choice use would also open avenues to controlling the axis orientation of the resulting \BTO~film which is key in electro-optic device fabrication. 
 
\section{Experimental Methods}

\subsection{Thin film synthesis}

We used a Veeco GenXplor MBE chamber with a custom made SVT Technology Services and Solutions metal-organic system for thin film synthesis of SrO and \BTO~. Elemental barium and strontium (99.99\% from Sigma Aldrich) were evaporated from effusion cells. We used a liquid precursor of titanium tetraisopropoxide (TTIP) (99.999\% from Sigma Aldrich) as a titanium and oxygen source. We also supplied additional molecular oxygen. Before thin film synthesis, we cleaned (100)-oriented {SrTiO$_3$} substrates with Micro 90 alkaline cleaner, acetone, isopropyl alcohol and deionized water. The substrates were then annealed in air at 1000 °C for 4 hours to obtain smooth step terraces.

Barium and strontium fluxes were first calibrated using a quartz crystal microbalance (QCM) with fluxes of $\sim 3 \times 10^{13}$ atoms/cm$^2$$\cdot$ sec. We first synthesized SrO sacrificial layers of 5-10 nm thickness by opening the strontium shutter in the presence of molecular oxygen at a pressure of $2\times 10^{-7}$ Torr and a substrate temperature of 500\degree C as measured by a thermocouple. We use an SrO sacrificial layer as it is a binary oxide that can be synthesized without extensive calibration by MBE and can dissolve in a matter of minutes in deionized water \cite{varshney_hybrid_2024}. SrO and \BTO~have a relatively large lattice mismatch ($\approx{9.6}$$\%$), but when SrO is used as a sacrificial layer  \BTO~grows epitaxially, enabling the deposition of high quality thin films.  In contrast, BaO has a better lattice match to \BTO~but requires intermediate buffer layers of {SrTiO$_3$} due to the interdiffusion of BaO and BaTiO$_3$ \cite{takahashi_sacrificial_2020}. 

The TTIP liquid precursor was heated to temperatures between 47\degree C - 60\degree C. We identify a growth window for stoichiometric film deposition with TTIP foreline pressures of 1.380 Torr to 1.771 Torr. The substrate and SrO film was heated to 800\degree C prior to the BaTiO$_3$ thin film deposition. We finally deposited 100 nm thick \BTO~by opening the barium shutters and inserting TTIP into the chamber using a heated gas cracking nozzle in the presence of oxygen at a pressure of $1\times 10^{-6}$ Torr.

\subsection{Structural and surface characterization}

X-ray diffraction (XRD) patterns were obtained using a Malvern PANalytical Empyrean diffractometer with Cu K$\alpha_1$ radiation 
($\lambda$ = 1.5406 $\si{\angstrom}$). X-ray reflectivity measurements were taken with the same tool to determine the film thickness.  We extracted the out-of-plane ($c$-axis) lattice parameters of our thin films using Nelson-Riley fits from XRD peak positions of \BTO~ \cite{nelson_experimental_1945}. Tapping mode atomic-force microscopy (AFM) was performed using an Asylum MFP-3D Origin+ instrument to determine the surface morphology of the thin films. 
We used a Zeiss Gemini 560 field emission scanning electron microscope (FE-SEM) to capture SEM images. 

\subsection{Thermocompression bonding}

We deposited 2 $\mu$m {SiO$_2$}  on the \BTO~thin films and on (100)-orientated silicon wafer substrates using Oxford Instruments' inductively coupled plasma chemical vapor deposition (Plasma Pro 100 ICP CVD) tool. {SiO$_2$} was deposited at 80°C. We then deposited a 15 nm chromium adhesion layer before depositing 100 nm gold on both the  {SiO$_2$} coated films and substrates using a PVD Products electron beam evaporation system. 

We used a Finetech fineplacer sigma advanced sub-micron bonder for thermocompression bonding of the gold coated \BTO~stack and the silicon substrate stack. Using both force and temperature in our bonding reduces the temperature requirement for bonding (unlike oxide-oxide bonding, which would require higher temperatures), which is critical when working with oxides like \BTO~that are prone to cracking \cite{tsau_fabrication_2003}. We varied the force, temperature and bonding times to optimize our bonding conditions (See Supplementary Figure S3). We found that the ideal bonding time for our best samples (with above 1 mm, crack-free transferred \BTO) to be 10 minutes, with a force of 1000 N and a temperature of 220°C. 

\subsection{Chemical lift-off}

The bonded stack was clamped with clips and immersed in room temperature DI water. After 12 hours, the SrO was completely dissolved, with  \BTO~on the silicon substrate. We left the stack to air dry. We found that using DI water at lower temperatures generally leads to a slower dissolve time, which helps reduce wrinkling \cite{varshney_hybrid_2024}. 
\vspace{6pt} 


\textbf{Author contributions:}
Synthesis of films by hybrid metal-organic MBE was performed by T.S.Z, with assistance from C.M.B, L.B.L, B.F.Y and J.A.M.  XRD characterization was performed by T.S.Z and C.M.B. AFM was performed by T.S.Z. RSM was performed by T.S.Z with assistance from K.Y.B. Fabrication was performed by T.S.Z with assistance from K.P and N.S. Etch tests were performed by A.M.D and C.Z. T.S.Z and J.A.M. wrote the paper, with input from all authors. J.A.M, D.R.B and M.L conceived of and guided the study.

\textbf{Acknowledgments:} We acknowledge useful conversations with Mac Hathaway, Arthur McClelland, Timothy Cavanaugh, Jason Tresback, William L. Wilson, Tomas Kraay. We acknowledge feedback on the manuscript from Suzanne Smith and Jenny Hoffman. This work was supported by the Center for Quantum Networks (Grant No. EEC-1941583) and the NSF under Grant. No.  OMA-2137723. Fabrication and electron microscopy was performed at Harvard University's Center for Nanoscale Systems (CNS), a member of the National Nanotechnology Coordinated Infrastructure Network (NNCI), supported by the National Science Foundation under NSF Grant No. 1541959. T.S.Z acknowledges support from the Harvard Center for Nanoscale system as the work was done when she was a center for nanoscale systems scholar. T.S.Z acknowledges the Harvard Quantum Initiative Post Baccalaureate programs' support as part of the work was done when she was a scholar. D.B. acknowledges financial support by the Intelligence Community Postdoctoral Fellowship.  A.M.D. acknowledges financial support from an HQI fellowship.

\textbf{Competing interests:} L.B.L is currently involved in developing barium titanate technologies at Epitactic Inc.  

\section*{Additional Information}

\textbf{Supplementary information is available at $\textcolor{orange}{XXX}$}

\clearpage
\bibliography{references}
\clearpage
\newpage

\widetext
\begin{center}
\textbf{\large Supplemental Materials: Transfer of Thin Film BaTiO$_3$ onto Silicon for Device Fabrication}
\end{center}

\setcounter{equation}{0}
\setcounter{figure}{0}
\setcounter{table}{0}
\setcounter{page}{1}
\setcounter{section}{0}
\makeatletter
\renewcommand{\theequation}{S\arabic{equation}}
\renewcommand{\thefigure}{S\arabic{figure}}
\renewcommand{\bibnumfmt}[1]{[S#1]}
\renewcommand{\thesection}{S\arabic{section}}

\begin{figure}[h]
    \includegraphics[width = \columnwidth]{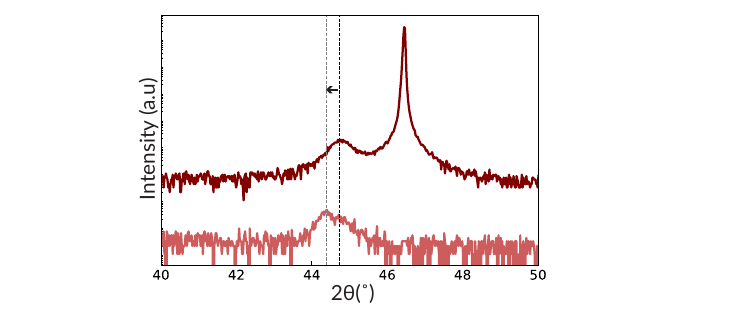}
    \caption{\textit{Peak Shifts after transfer of \BTO~onto silicon}. The zoomed in XRD of the \BTO~002 peak shows a big peak shift after transfer onto silicon. The lattice parameter increased from 4.057 $\si{\angstrom}$ to 4.079 $\si{\angstrom}$,  indicating an in-plane compressive strain. Compared to our films that did not show such large peak shifts, we hypothesize that this shift is mainly due to the change in ramp rates for thermocompression bonding, which affected the thermal expansions of the different materials. In this film, we used a force ramp rate of 100 N/s and a temperature ramp rate of 20\degree C/s while in the films that did not have drastic peak shifts we lowered our ramp rates to a force ramp rate of 6 N/s and a temperature ramp rate of 1\degree C/s. Using slower ramp rates should not only help with reducing peak shifts but in thicker films this will also avoid instances of the thin film cracking as we cycle through its Curie temperature of 130\degree C. One component we have not accounted for is that this was a much thinner film (30 nm) compared to later films that did not have larger peak shifts (100 nm). So it could be possible that the thickness also played an important role in peak shifts.}
     \label{fig:peak shifts bonding}
\end{figure}

\begin{figure}
    \includegraphics[width = \columnwidth]{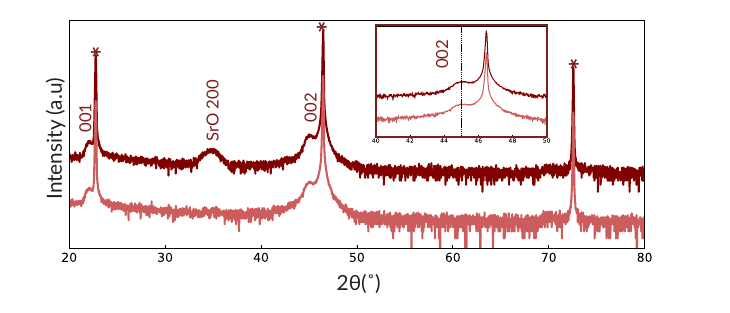}
\caption{\textit{SrO degradation in air}. The XRD of \BTO~on SrO post synthesis (maroon) and after the SrO degraded in air (pink). In the insert we demonstrate that the 002 peak still shifts due to the SrO layer being gone. We obtain a c-axis lattice parameter shift from 4.017 $\si{\angstrom}$ to 4.022 $\si{\angstrom}$ indicating similar compressive in-plane strain. In this case  the \BTO~film could be more strained to the {SrTiO$_3$} substrate with the SrO layer gone. Another observation we made in this example is that the SrO layer was 22 nm and the \BTO~was 12 nm, which led to the SrO degrading in air. In our other samples, we generally noted that having a thicker SrO layer than \BTO~led to the SrO degrading much quicker in air. It would potentially be helpful to make this buffer layer thinner than the \BTO~thin film to avoid degradation of SrO.}
     \label{fig:peak shifts SrO}
\end{figure}

\begin{figure}
    \includegraphics[width = \columnwidth]{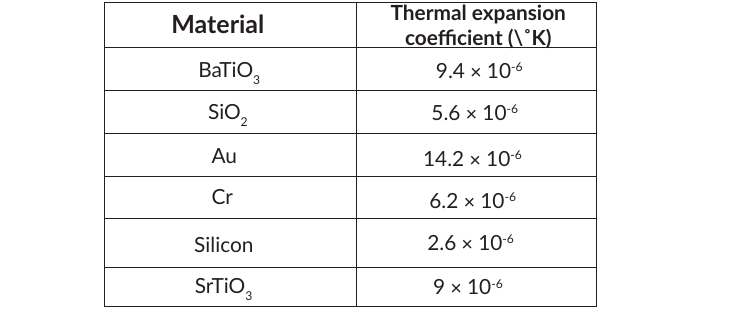}
    \caption{\textit{Thermal expansion coefficients}. The different thermal expansion coefficients of the materials in our bonding stack. Gold will have the largest thermal expansion for each change in \degree K, meaning that this will lead to some strain induced on the \BTO~thin film. Using a bonding stack with less of such differences may help reduce instances of inducing strain on the \BTO~transferred layer \cite{mse_supplies_llc_nd_list_nodate}.}
    \label{fig:thermal expansions}
\end{figure}

\begin{figure}
    \includegraphics[width = \columnwidth]{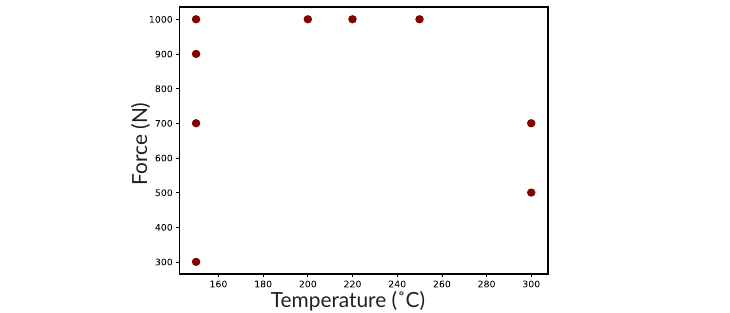}
    \caption{\textit{Bonding temperatures and forces}. The different bonding conditions we explored. We found the ideal conditions to be a bonding force of 1000 N and a temperature of 220 \degree C. With thicker films we observed that having a bonding temperature of anywhere between 220 \degree C and 300 \degree C still gave good bonds at the same force of 1000 N. In this plot we have not shown time, bonding times below 5 minutes tended to be weak and above 10 minutes there was not much improvement in the bonded films. Additionally slow ramp times are important in maintaining the as synthesized tetragonal structure of our films, so we used the slowest times available to us on the tool for both ramp up and ramp down. }
    \label{fig:bonding conditions}
\end{figure}

\begin{figure}
    \includegraphics[width = \columnwidth]{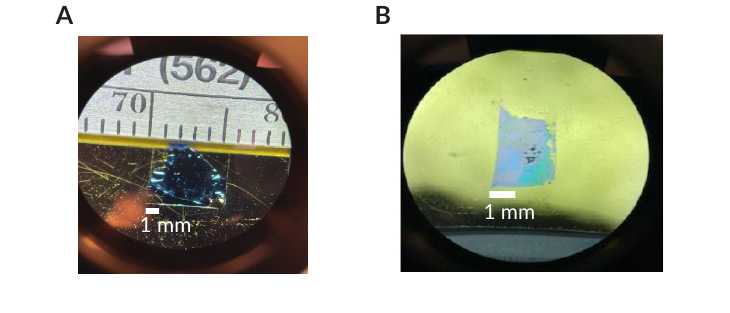}
\caption{\textit{\BTO~on silicon images}. \textbf{A.} Optical image of \BTO~transferred onto silicon. The ~100 nm thickness leads to the blue color. \textbf{B.} Optical image of a thinner film, about 30 nm \BTO~transferred onto silicon. In the thinner films, we had challenges differentiating between the \BTO~and SiO$_2$ layer as it has a color gradient (different colors because the SiO$_2$ give multiple colors) but a very thin film of \BTO~was transparent.}
    \label{fig:optical images}
\end{figure}

\begin{figure}
    \includegraphics[width = \columnwidth]{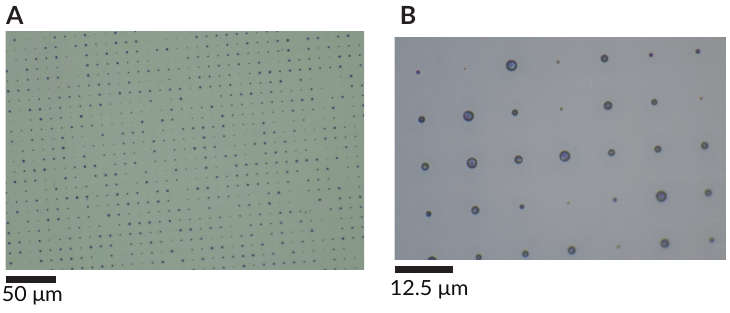}
    \caption{\textit{Etched nano-pillars images}. \textbf{A.} The optical image of \BTO~nano-pillars at 20X magnitude.\textbf{B.} The optical image of the same nano-pillars at 100X magnitude.}
    \label{fig:Etched nanopillars images}
\end{figure}

\end{document}